\documentclass[aps,reprint,superscriptaddress,floatfix]{revtex4-2}

\usepackage{graphicx,amssymb,amsmath,xcolor}
\usepackage[normalem]{ulem}
\graphicspath{{figs/}}

\usepackage{color}
\usepackage{colortbl}

\begin{document}

\title{Constraints on primordial black holes 
from observation of stars in dwarf galaxies}


\author{Nicolas Esser}
\affiliation{Service de Physique Th\'{e}orique, Universit\'{e} Libre
de Bruxelles (ULB),\\CP225 Boulevard du Triomphe, B-1050 Bruxelles,
Belgium}
\author{Peter Tinyakov}
\affiliation{Service de Physique Th\'{e}orique, Universit\'{e} Libre
de Bruxelles (ULB),\\CP225 Boulevard du Triomphe, B-1050 Bruxelles,
Belgium}

\date{\today}

\begin{abstract}
We propose a way to constrain the primordial black hole (PBH) abundance in the
range of PBH masses $m$ around $10^{20}$g based on their capture by Sun-like
stars in dwarf galaxies, with subsequent star destruction. We calculate
numerically the probability of a PBH capture by a star at the time of its
formation in an environment typical of dwarf galaxies. 
Requiring that no more
than a fraction $\xi$ of stars in a dwarf galaxy is destroyed by PBHs
translates into an upper limit on the PBH abundance. For the parameters of
Triangulum II and $\xi=0.5$, we find that no more than $\sim 35\%$ of dark matter can
consist of PBHs in the mass range $10^{18} - \text{(a few)}\times
10^{21}$g. The constraints depend strongly on the parameter $\xi$ and may significantly improve if smaller values of $\xi$ are established from observations. An accurate determination of $\xi$ from dwarf galaxy modeling is thus of major importance.
\end{abstract}



\maketitle

\section{Introduction}

As the parameter space of particle dark matter (DM) models becomes more and
more constrained by the direct and indirect detection experiments
\cite{Arcadi:2017kky,Roszkowski:2017nbc}, other nonparticle candidates start
attracting considerable attention. Among these candidates, an interesting
possibility is that DM is composed of primordial black holes (PBHs), which
could have been created in the early Universe and survived until today
\cite{Zeldovich:1967lct,Hawking:1971ei}. A variety of mechanisms to create
such black holes has been proposed in the literature (see
Ref. \cite{Carr:2020gox} for a review). Their masses $m$ may range from $\sim
10^{17}$~g, below which they would evaporate too fast by Hawking radiation, to
tens of solar masses where they could contribute to, or potentially explain, 
the black hole merger events observed by gravitational wave detectors
\cite{LIGOScientific:2018mvr,LIGOScientific:2020ibl}. Their relative abundance
$f = \Omega_{\rm PBH}/\Omega_{\rm DM}$ can typically be adjusted to match the
whole or a fraction of the observable DM.

For PBH masses roughly above $\sim 10^{23}$g, their contribution to the total
DM density has been constrained from various observations \cite{Carr:2020gox}.
However, a large --- several orders of magnitude --- range of masses around
$10^{20}$g remains virtually unconstrained. These asteroid-mass PBHs are
particularly difficult to constrain as they have atomic size, and are too few
to hope for a direct detection. A proposal has been made to use the capture of
these PBHs by neutron stars (NS) and white dwarfs in DM-rich environments like
dwarf galaxies \cite{Capela:2013yf,Capela:2012jz,Capela:2014ita}; however, no
observations of NS in dwarf galaxies exist at present.

In this paper we propose a way to constrain the abundance of asteroid-mass PBHs in the mass range
around $10^{20}$g based on their capture by {\em main sequence}, {\em sun-like} stars. While
main sequence stars offer less favorable conditions for capture as compared to
NS and white dwarfs, the advantage is that they are routinely observed in
dwarf galaxies. The direct PBH capture during the star lifetime is not efficient in the case
of main sequence stars \cite{Montero-Camacho:2019jte}; we will concentrate
instead on their capture at the stage of star formation, which is the dominant mechanism for main sequence stars, as will be explained below.

If a Sun-like star captures a PBH of  mass around $m\sim 10^{20}$g, the
latter starts accreting the star matter. Assuming the Bondi accretion, which
is justified by the fact that the Bondi radius $r_B = 2Gm/c_s^2 \sim 5\times
10^{-3}$cm is much larger than the interatomic distance, the accretion time is
estimated as
\begin{equation}
t_{\rm acc} = \frac{c_s^3}{4\pi\rho_\star G^2m} 
=  5.2\times 10^6 {\rm yr} \left(10^{20}{\rm g} \over m\right),
\label{eq:destruction}
\end{equation}
where we have used the value of the sound speed in the Sun
core $c_s = 511$~km/s and the core density $\rho_* = 146$~g/cm$^3$. This is
much smaller than the star lifetime $t_* \sim 10^{10}$~yr, so that a star
that has captured a PBH gets rapidly destroyed. Mere observations of stars
then constrain the probability of capture, which translates into constraints
on the PBH abundance $f$. 

The rest of this paper is organized as follows. In Sec.~\ref{sec:capture-pbh-main}
we calculate the probability of the PBH
capture by a star in a DM halo with reference values of parameters. In
Sec.~\ref{sec:constr-from-observ} we show how this result can be used to constrain
the PBH abundance in concrete dwarf galaxies. Section~\ref{sec:conclusion}
contains concluding remarks.

\section{Capture of PBHs by main sequence stars}
\label{sec:capture-pbh-main}

For a successful capture a PBH must first get on a bound orbit that crosses
the star, and then gradually loses energy during periodic collisions with the
star until it finally sinks completely inside the latter. The second,
``cooling'' stage is common to both capture mechanisms (the direct capture during the star lifetime and the capture at the time of star formation), so let us first  
look at this stage in more quantitative terms. 

A PBH crossing the star loses energy to the dynamical friction
\cite{Chandrasekhar:1943ys} described by the force
\begin{equation}
F = - 4\pi G^2 m^2 \rho_* {\ln \Lambda\over v^2}
\label{eq:DF}
\end{equation}
where $m$ is the PBH mass, $\rho_*$ the mean star density, $v$ the PBH velocity,
and $\ln \Lambda\approx 30$  the Coulomb logarithm.  This mechanism is very
inefficient. In addition, most of the PBHs that eventually get captured start
on extremely elongated orbits and spend most of their time in a loss-free
Keplerian  motion. Everywhere except in the vicinity of the star, such orbits are
well approximated as radial and are characterized by their apastron $r_{\rm
  max}\gg R_*$, $R_*$ being the star radius. The initial value of $r_{\rm
  max}$ together with Eq.~(\ref{eq:DF}) determine the duration of the
cooling stage. Following the method of Ref.~\cite{Kouvaris:2010jy}, the cooling time can be estimated as 
\begin{equation}
t_{\rm cool} \sim {\pi M_* R_*\over m \ln \Lambda} 
\sqrt{{r_{\rm max}\over R_g}}\sim 10^{10}\, {\rm yr} 
\left({r_{\rm max}\over 100 {\rm AU}}\right)^{1/2}
\left({10^{20}{\rm g} \over m}\right)
\label{eq:t_cool_estim}
\end{equation}
where $M_*$ is the star mass and $R_g=2GM_*$ its gravitational radius. The PBHs
that have cooling times exceeding $\sim 10^{10}$~yr do not get captured.

Apart from the insufficient time, the cooling may be unsuccessful if
interrupted by the deviation of the PBH from the star-crossing orbit due to
perturbations produced by, e.g., nearby stars. Such PBHs stop losing energy
and do not get captured. To quantify the effect of perturbers we first note
that for very extended, nearly radial orbits the periastron $r_{\rm min}$ is
determined by the PBH angular momentum $J$ with respect to the star,
\begin{equation}
r_{\rm min} = {J^2/ (m^2 R_g)}.
\label{eq:rmin}
\end{equation}
The condition under which the orbit crosses the star $r_{\rm min}<R_*$ translates
into the maximum angular momentum, $J/m<J_{\rm max}/m = \sqrt{R_* R_g}$. In
the presence of perturbations the angular momentum may change; successful
cooling requires that these changes are smaller than $J_{\rm max}$. For the
estimate, assume the original radial trajectory $x(t)$ is perturbed in the
plane $(x,y)$ by a small potential $U(\vec r)$. The change of the angular
momentum over the time of one free fall from $r_{\rm max}\gg R_*$ to $r\sim R_*$ is
\[
\Delta J / m = \int _0^{T/4} x^2(t) U_{xy}dt,
\]
where $T$ is the period of the Keplerian orbit and $U_{xy} = \partial_x \partial_y U$.  
Substituting the unperturbed motion and neglecting the variations of $U_{xy}$
along the trajectory we get 
\begin{equation}
\Delta J / m = {5\pi\over 16}{r_{\rm max}^{7/2}\over R_g^{1/2}} U_{xy}. 
\label{eq:DeltaJ}
\end{equation}
In the case when the perturbation is caused by another star, considered
static,  at a distance $d\gg r_{\rm max}$, we obtain from Eqs.~(\ref{eq:rmin}) and (\ref{eq:DeltaJ}) the condition on the periastron of the perturbed orbit
\begin{equation}
r_{\rm min} = \alpha\,  r_{\rm max} \left({r_{\rm max}\over d }\right)^6 <R_*,
\label{eq:constraint2}
\end{equation}
where $\alpha$ is a calculable numerical coefficient of order 1 depending on the
direction to the perturber. This imposes a second constraint on $r_{\rm
  max}$. 
While the first condition re-
sulting from Eq.~(\ref{eq:t_cool_estim}) becomes less important as the PBH
mass increases, the second one, Eq. (6), is independent of
the mass. We will use these constraints later when calculating the capture probability.

Let us now turn to the first stage of capture which sets initial conditions
for the cooling. Two mechanisms have been considered in the literature:
capture during the star lifetime, and at the star formation. In the first
case, during the star lifetime some PBHs from the ambient DM halo may collide
with the star, lose energy due to dynamical friction,
Eq.~(\ref{eq:DF}), and become gravitationally bound. In the context of capture by neutron stars this process has been considered in Ref.~\cite{Capela:2012jz}. 
Only very slow PBHs can become bound after a single
collision. Their energies then become of order $-E_{\rm loss}$,
which determines the initial size of their orbits. In the case of main
sequence stars the energy loss is very small and is estimated from
Eq.~(\ref{eq:DF}) as \cite{Capela:2013yf}
\begin{equation}
E_{\rm loss} \sim {2Gm^2\over R_*} \ln \Lambda. 
\label{eq:avEloss}
\end{equation}
Converting this to the orbit size we find
\[
r_{\rm max} = {M_* R_*\over 2 m \ln\Lambda} = 7.5\,\text{kpc} 
\left({10^{20}\text{g}\over m}\right).
\] 
This is by many orders of magnitude larger than needed to satisfy any of the
two cooling constraints, cf. Eqs.~(\ref{eq:t_cool_estim}) and (\ref{eq:constraint2}) . We conclude, in agreement with 
Ref.~\cite{Montero-Camacho:2019jte}, that
the direct capture does not work in the case of main sequence stars.

Consider finally the capture at the star formation
\cite{Capela:2012jz,Capela:2014ita}. When the star is formed from a gas cloud,
the contracting baryons create a time-dependent gravitational potential that
drags along the DM particles (PBHs in this case) which lose energy and develop
a cuspy density profile centered on the star. Only a fraction of PBHs that are gravitationally bound to the cloud and repeatedly cross it during the contraction are significantly affected. By the end of the contraction some of
them settle on orbits crossing the newly formed star, and thus enter the
cooling stage.

We now proceed to the calculation of the number of
PBHs captured by a star in this way. The capture is a random process;
the number of PBHs captured by a single star follows the Poisson
distribution characterized by the mean captured PBH mass. The latter can
be written as $f\bar M$, where $f$ is the PBH abundance and $\bar M$ is the captured mass assuming
all of the DM
consists of PBHs. For an ensemble of stars in identical
conditions the mean captured mass $f\bar M$ determines  the fraction $\xi$ of
stars that have been infected (and eventually destroyed)
by PBHs of mass $m$, $\xi = 1-\exp(-f\bar M /m)$. The fraction
of destroyed stars $\xi$ is assumed to be constrained from observations. Inverting 
the above relation we thus obtain the constraint on the PBH abundance $f$,
\begin{equation}
f < { m \over\bar M} \ln {1\over 1-\xi}.
\label{eq:M-of-xi}
\end{equation}
Note that no constraints arise when the rhs of Eq.~(\ref{eq:M-of-xi}) is
larger than one as $f\leq 1$ by definition.  
 
Following the logic of Refs. \cite{Capela:2012jz,Capela:2014ita}, we calculate
the mean captured mass $\bar M$ by numerically evolving randomly generated PBH
trajectories, one at a time, in the time-dependent gravitational field of a
contracting baryon cloud, counting those that correspond to star-crossing
orbits at the end of the contraction and satisfy the cooling
conditions. Knowing the DM abundance and distribution, and the fraction of
``successful'' trajectories, we can determine $\bar M$. The detailed steps of
the simulation are as follows.

We assume that the initial distribution of baryons is a uniform sphere of size
$R_C = 4300$~AU and density $1.78\times 10^{-18}$~g/cm$^3$ as corresponds to
the parameters of a pre-stellar core of a star of $1\,M_\odot$
\cite{Kirk:2005ng}. The baryonic sphere is slowly contracted in size, while
its profile is gradually changed from a uniform one to the actual star density
profile rescaled to the current sphere size, with
the total mass being constant.
At the end of the contraction the result is a star of Sun radius $R_*=R_\odot$ and with the Sun density profile \cite{2000A&A...358..593S}. 
In the adiabatic approximation the
precise way of contraction does not matter; we use the linear decrease in
radius as well as a linear evolution of the profile. We have checked that the results are stable with respect to changes of the contraction law provided it is much slower than the free fall (in practice, several times slower is sufficient \cite{Capela:2012jz}).

The initial conditions for the PBH trajectories sample the DM distribution. We assume that the DM is uniformly distributed in space with density $\rho_{\rm DM}$ and follows the Maxwell velocity distribution with the dispersion $\bar v$. 
The reference values $\rho_{\rm DM}=100$~GeV/cm$^3$ and $\bar v = 7$~km/s
are adopted at this stage and will be rescaled in 
Sec.~\ref{sec:constr-from-observ} according to conditions
in concrete dwarf galaxies. 

In position space, we sampled the
region $r< 20 R_C$. We have checked that increasing further the size of this
region does not change the results as those trajectories that start further
away do not satisfy the cooling conditions and will in any case be rejected.

In velocity space, we limited the sampled region to $v< v_{\rm esc}$, where
$v_{\rm esc}= \sqrt{(3GM_\odot/R_C)}=0.79\text{km/s}$ is the escape velocity
from the center of the uncontracted cloud. Note that, since $v_{\rm esc}\ll
\bar v$, 
the volume of this region is proportional to $v_{\rm esc}^3$. To increase the computational
efficiency we further limited the sampled region to small tangential
velocities $v_\perp < \sqrt{2GM_\odot R_\odot}/r$; this condition eliminates
trajectories with large angular momenta since they do not cross the star 
after the contraction of the cloud.
The volume of the resulting phase space, as well as the total DM mass
it contains, was calculated analytically. 

Not all the initial conditions from the above region of phase space correspond
to bound particles. We sampled the whole region randomly and discarded the
unbound (positive energy) trajectories keeping track of their fraction. The
remaining ones were evolved in the gravitational field of the contracting
cloud; those that did not have periastrons within the star radius at the end of the contraction (a very
small faction) were again discarded. 
Combining all the fractions, the total amount of the DM sampled by
the accepted trajectories is $8.34\times10^{21}$g. We have simulated in this
way a total of $4\times10^{6}$ trajectories, recording for each of them the
periastron and the apastron after the contraction.

To compute the mass of PBHs that actually get captured, the cooling conditions
have to be checked for each trajectory. The fraction of trajectories
respecting both these conditions, multiplied by the total mass previously
obtained, gives the mean mass $\bar M$ of PBHs captured by the star.  The
estimate (\ref{eq:t_cool_estim}) cannot be used to check individual
trajectories as it was based on an average energy loss. Instead, we use Eq.~(\ref{eq:DF}) directly to calculate the cooling time for each trajectory,
taking into account its parameters and making use of the actual star density
profile \cite{2000A&A...358..593S}. The problem is, in general, complicated
since the energy and angular momentum losses have to be calculated every time
the PBH crosses the star, all the way until it sinks inside. However, several
approximations may be used to simplify the task.

Clearly, the cooling time is dominated by the first stages when the orbits are
nearly radial. One may check that for such orbits the star crossing episodes
have negligible effect on the periastron $r_{\rm min}$ and mainly affect the
apastron $r_{\rm max}$. Moreover, the energy loss at one crossing is
practically independent of $r_{\rm max}$ and only depends on $r_{\rm
  min}$. Since the motion between the crossings is Keplerian, one can
calculate the cooling time analytically in terms of $E_{\rm loss}(r_{\rm
  min})$. Requiring that the cooling time does not exceed the star lifetime
$t_*=10^{10}$~yr, one obtains the constraint on $r_{\rm max}$,
\begin{equation}
r_{\rm max} < {E_{\rm loss}^2(r_{\rm min}) \over 2 \pi^2 G M_* m^2} t_*^2 .
\label{eq:rmax-constraint}
\end{equation}
Note that if we use here the estimate (\ref{eq:avEloss}) for $E_{\rm loss}$ instead 
of $E_{\rm loss}(r_{\rm  min})$,  we recover Eq.~(\ref{eq:t_cool_estim}).
The function $E_{\rm loss}(r_{\rm min})$ was calculated numerically for the
PBH mass $m=10^{20}$~g. For other PBH masses it can be obtained by a simple
rescaling $E_{\rm loss}\propto m^2$, cf. Eq.~(\ref{eq:avEloss}). We have
checked that the analytically calculated cooling time is well reproduced by a
direct numerical integration for several trajectories.

The second cooling constraint arises from the requirement that the PBH
trajectory is not deviated from the star-crossing regime by gravitational
perturbations. We model these perturbations by static randomly placed stars
with the density $n_*$. Making use of the three-dimensional version of
Eq.~(\ref{eq:DeltaJ}), we determine the maximum size of the trajectory such
that the deviations remain small in at least 50\% of random realizations of
the perturbers. We discard the trajectories that have larger apastrons.  Note
that this second condition is independent of the PBH mass.

\begin{figure}[th]
\includegraphics[width=1.\columnwidth]{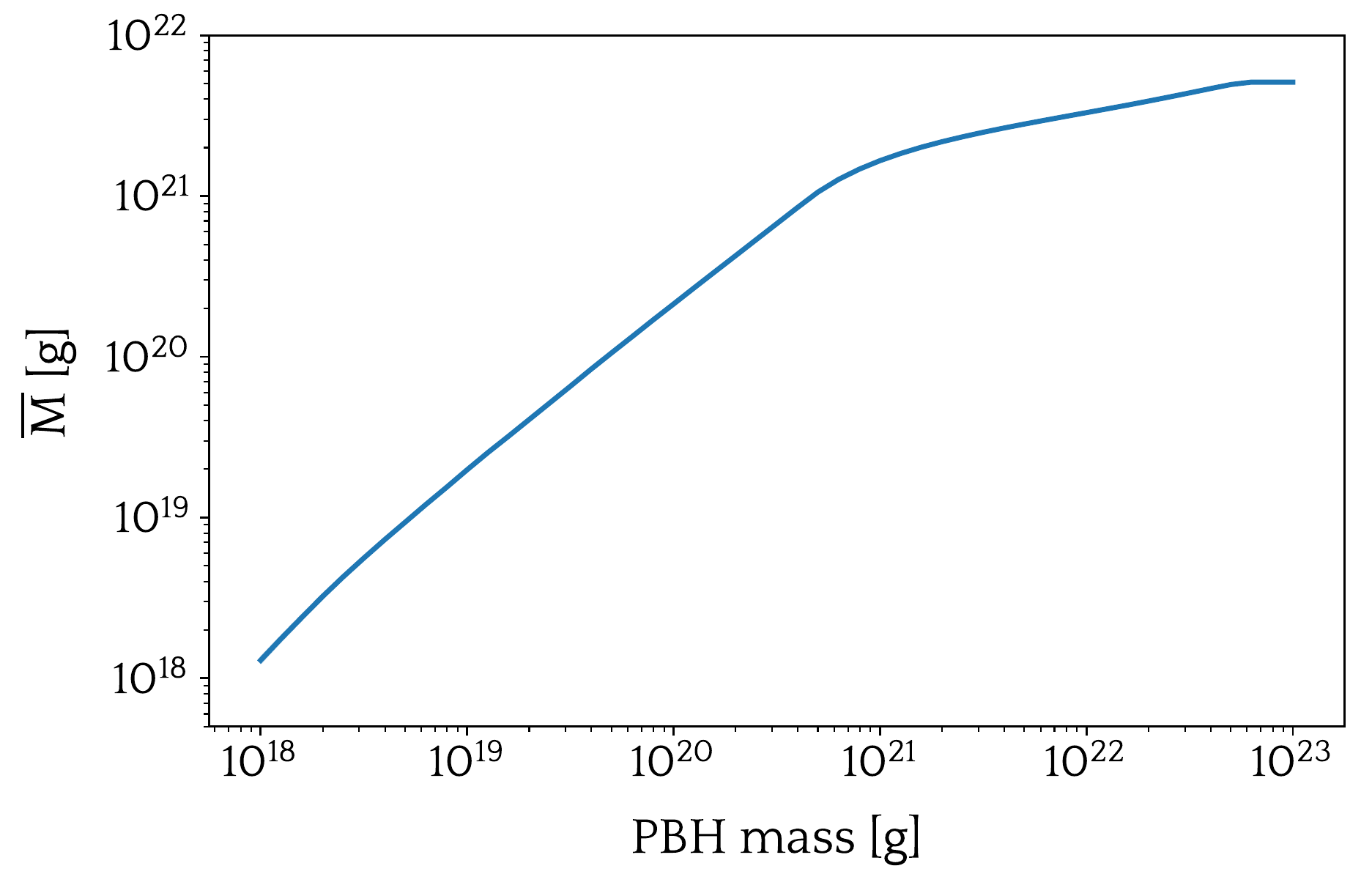}
\caption{\label{fig:meanM}
Mean captured mass $\bar M$ as a function of the PBH mass $m$ for the star
density $n_* = 9.2\times 10^{-3}{\rm pc}^{-3}$ and reference values $\rho_{\rm
  DM} = 100$GeV/cm$^3$ and $\bar v = 7$km/s. The statistical uncertainty of
the simulation is not visible on the plot.}
\end{figure}
The net result of the capture process is summarized in Fig.~\ref{fig:meanM},
which shows the mean captured mass $\bar M$ as a function of the PBH mass
$m$. The star density of $n_* = 9.2\times 10^{-3}{\rm pc}^{-3}$ was used in producing this figure, which is about the highest among the dwarf galaxies
considered in the next section. The behavior is approximately linear below
$m=\text{(a few)}\times 10^{20}$g; the cooling time constraint is dominant in
this region. At higher masses the constraint due to perturbers starts to
dominate and the behavior flattens out.

\section{Application to observed dwarf galaxies}
\label{sec:constr-from-observ}

The mean captured mass $\bar M$ together with Eq.~(\ref{eq:M-of-xi}) allow one
to constrain the PBH abundance $f$ given the value of $\xi$. 
Since the captured mass is directly proportional to the factor $\rho_{\rm
  DM} /{\bar v}^3$, the strongest constraints come from the environments with the largest DM density and smallest velocity dispersion, the conditions realized in the ultra-faint dwarf galaxies. Several tens of dwarf galaxies have been
observed around the Milky Way; for a review see
Ref.~\cite{Simon:2019nxf}. Table~\ref{tab:1} lists the essential properties of
five of the most promising ones.
\begin{table}[ht]
\begin{tabular}{|c|c|c|c|c|c|}
\arrayrulecolor{gray}\hline
 & $R_{1/2}$ & $\sigma$ & $\rho_{\rm DM}$ & $n_*$ & $\eta$ \\
 & [pc]  & [km/s] & $[\text{GeV/cm}^3$] & $[10^{-3} \text{pc}^{-3}]$ 
&
\\ \hline
Triangulum II & $16$ & $<5.9$ & $161$ & $9.2$ & $0.95$ \\ \hline
Tucana III & $37$ & $<2.1$ & $3.7$ & $0.67$ & $0.51$ \\ \hline
Draco II & $19$ & $<10.2$ & $343$ & $2.6$ &$0.39$ \\ \hline
Segue 1 & $24$ & $6.4$ & $85$ & $2.1$ &$0.39$ \\ \hline
Grus I & $28$ & $5.0$ & $38$ & $9.6$ & $0.37$ \\ \hline
\end{tabular}
\caption{\label{tab:1}
The parameters of dwarf galaxies.}
\end{table}

For each galaxy, the directly measured parameters are the line-of-sight
velocity dispersion of stars that determines the 3D dispersion $\sigma$, and
the projected half-light radius $R_{1/2}$ related to the 3D one by $R_{3D} =
4/3 R_{1/2}$. We assume that the velocity dispersions of stars and of DM are
the same, and that both follow the Maxwell distribution. We then obtain the DM
dispersion in the star rest frame $\bar v = \sqrt{2} \sigma$. Knowing 
$R_{1/2}$ and $\sigma$ we can calculate the mass $M_{1/2}$ within the
half-light radius by means of Eq. (1) of
Ref.~\cite{Simon:2019nxf}. Assuming baryons are subdominant, from $M_{1/2}$
and $R_{1/2}$ one obtains the DM density $\rho_{\rm DM}$.  Knowing the
luminosity one can then calculate the number density of stars $n_*$ within
$R_{3D}$ and check that their mass density is indeed subdominant to that of
the DM.  We have therefore all the parameters needed to determine  the mean
captured mass $\bar M$ as a function of $m$ in the conditions of a
concrete dwarf galaxy.

The constraining power of a given dwarf galaxy is proportional to the ``merit
factor''
\[
\eta = {\rho_{\rm DM}\over 100~
\text{GeV}/\text{cm}^3} 
\left({7\text{km/s}\over \sqrt{2} \sigma}\right)^3.
\]
The merit factors of observed dwarf galaxies are listed in Table \ref{tab:1}.  Where
only the upper limit on $\sigma$ exists, this limit was used in
calculations, which is conservative since, including the factor $\sigma^2$ coming from
the DM density $\rho_{\rm DM}$, the merit factor is proportional to
$1/\sigma$. The highest merit factor corresponding to the strongest constraints is
found for Triangulum II. It is important to note, however, that this galaxy is
not an exception; several other galaxies have similar constraining power.

Once the mean captured mass per star, $\bar M$, is determined for a given 
galaxy, one may use  Eq.~(\ref{eq:M-of-xi}) to constrain the PBH abundance provided 
that the maximum allowed fraction of destroyed stars $\xi$ is known.  
While there exist quantitative models of dwarf galaxies, see
Ref.~\cite{Simon:2019nxf} for a review, no estimates of $\xi$ have yet been
performed. Such estimates require a dedicated analysis of galaxy evolution
models and lie beyond the scope of this paper. 
Nonetheless, including the new mechanism of star destruction by PBHs in galaxy 
evolution codes and requiring that the present-time properties (such as the stellar 
mass, stellar-to-halo mass ratio, mean metallicity, and metallicity dispersion) 
correspond to the ones observed in UFDs should allow one to determine the value of 
$\xi$. For now, we treat $\xi$ as a free parameter and choose $\xi =0.5$ as a benchmark value,
which corresponds to half of the stars in a galaxy, typically several
hundred, having been destroyed by PBHs. Since this would imply order 1 changes in the galaxy modeling, we expect that this
choice is reasonable. For illustration, we also present the constraints for
$\xi=0.2$. Note that the scaling of the constraints with $\xi$ trivially follows from Eq.~(\ref{eq:M-of-xi}).

\begin{figure}[ht]
\includegraphics[width=1.\columnwidth]{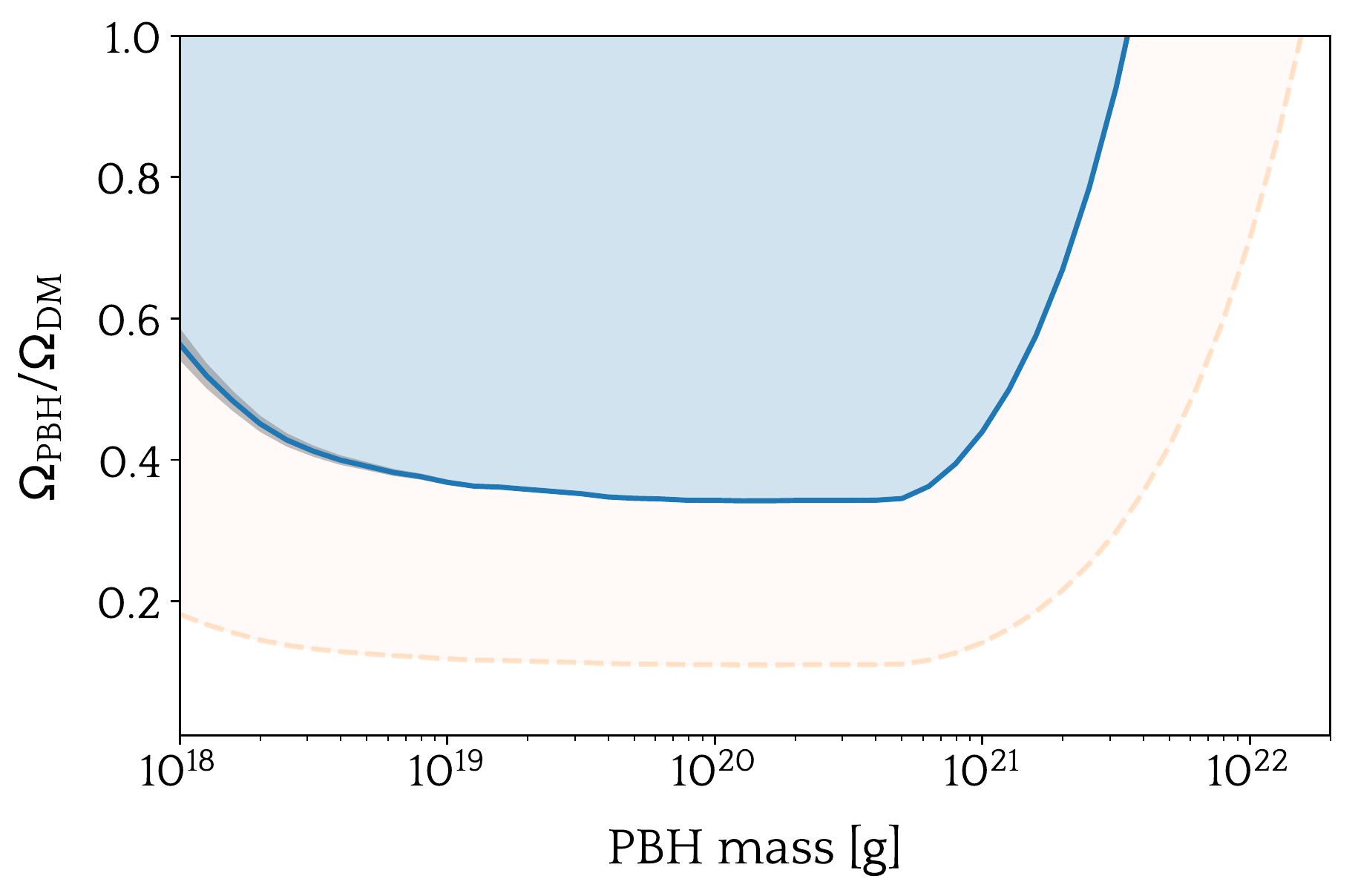}
\caption{\label{fig:constraints} Constraints on the PBH abundance $f =
  \Omega_{\rm PBH}/\Omega_{\rm DM}$ as a function of the PBH mass for the
  Triangulum II dwarf galaxy. Blue and orange regions are excluded assuming
  $\xi=0.5$ and $\xi=0.2$, respectively. The band, barely visible at low
  masses, represents the statistical uncertainty of the simulation (not shown
  for the orange curve).}
\end{figure}
In Fig.~\ref{fig:constraints} we show the resulting constraints on the PBH
abundance $f = \Omega_{\rm PBH}/\Omega_{\rm DM}$ as a function of the PBH mass
for Triangulum II, the most constraining galaxy of those listed in
Table.~\ref{tab:1}. The blue and orange regions are excluded assuming
$\xi=0.5$ and $\xi=0.2$, respectively. We do not show the constraints for
masses below $10^{18}$~g because in this range of masses the time of star
destruction, Eq.~(\ref{eq:destruction}), becomes comparable or exceeds the
star lifetime.

\section{Conclusion}
\label{sec:conclusion}

The most important conclusion from our results is that in 
DM-rich environments such as dwarf galaxies the probability of 
capture of an asteroid-mass PBH by an ordinary star may be of order 
1, provided these PBHs constitute a substantial fraction of the 
dark matter. This probability was thought to be negligible in the existing 
literature, the reason being that capture at the stage of star formation 
has not been included. While for neutron stars and white dwarfs this mechanism 
gives a contribution of the same order as the direct capture during the star
lifetime, it is by far dominant in the case of ordinary stars. 
This opens a possibility to constrain 
the PBH abundance in the mass range $10^{18} - 10^{22}$~g where no other
constraints exist at the moment. The advantage of the ordinary stars as 
compared to neutron stars and white dwarfs is that they {\em are} observed in dwarf galaxies. 

We performed a detailed calculation of the PBH capture by a star at its 
formation. Similar calculations have been done previously in the context 
of compact stars. We have applied these calculations to ordinary stars, including, in addition, the effect of gravitational perturbations arising from nearby stars, and 
taking into account the actual star density profile. 
The results of our computation are summarized in Fig.~\ref{fig:meanM} in terms 
of the mean captured mass per star $\bar M$ for the reference parameters of 
the ambient DM halo, $\rho_{\rm DM} = 100$GeV/cm$^3$ and $\bar v = 7$km/s. This part 
of our calculations does not depend on astrophysical uncertainties.

We then converted the mean captured mass into constraints on the PBH
abundance, assuming that no more than a fraction $\xi$ of all stars in a
dwarf galaxy has been destroyed by PBHs.  To derive the constraints for a
concrete dwarf galaxy we rescaled the $\bar M$ of Fig.~\ref{fig:meanM}
according to the galaxy merit factor, Table \ref{tab:1}, and took into account
the actual star density in the Galaxy, which determines precisely how the constraints
are cut off at high masses. We present the result for Triangulum II in
Fig.~\ref{fig:constraints}, using as a benchmark the value $\xi=0.5$.

The resulting constraints depend crucially on the assumed value of $\xi$: they are marginal for the adopted value $\xi=0.5$ and would disappear for $\xi \gtrsim 0.9$. Clearly, a quantitative determination of this parameter from dwarf galaxy modeling is of major importance. The example of $\xi=0.2$ in Fig.~\ref{fig:constraints} shows that limiting $\xi$ to lower values may significantly improve the constraints on the PBH abundance. 

The conversion of $\bar M$ into the constraints on $f$ involves a 
few additional astrophysical assumptions which may need to be better quantified together with a more precise determination of $\xi$. In the present calculation, we neglected the distribution of stars in masses assuming all stars in a dwarf galaxy have masses $\sim 1M_\odot$. We also neglected the variations of the DM density within a dwarf galaxy. In view of 
higher DM density, a cuspy profile is likely to strengthen the constraints.

Finally, a more accurate data on dwarf galaxies, notably the measurements of velocity dispersions for galaxies where only the upper bounds exist at present, may further strengthen the results. New data are expected in the near future from the ongoing and upcoming surveys \cite{Simon:2019nxf}. 

In this paper we have discussed only constraints resulting from
a mere disappearance of stars in dwarf galaxies. There may exist other, independent signatures of star conversion to black holes by captured PBHs. First, it is likely that last stages of accretion are catastrophic events with supernova-type energy release. If so, these events may be important in the energy and/or gas balance of a dwarf galaxy and will have to be taken into account in its modeling. Furthermore, with total energy release of order of a supernova, these events will be easily observable even if they last for millions of years. Their nonobservation in this case may result in additional 
constraints on $\xi$ and, therefore, on PBH. Second, for sizeable values of $\xi$ the dwarf galaxy will have a large population of subsolar mass black holes ---  remnants of star destructions. These black holes may potentially be observed, e.g., through gravitational waves produced in their coalescence. 
We leave these questions for future study.

\section*{Acknowledgements}

We are grateful to S.~Clesse and M.~Pshirkov for useful discussions and comments on the manuscript, and to J.~Simon and E.~Kirby for the discussion of dwarf galaxies. This work is supported in part by the Institut
Interuniversitaire des Sciences Nucléaires Grant No. 4.4503.15. N. E. is a FRIA grantee of the Fonds de
la Recherche Scientifique – FNRS.

%

\end{document}